\documentclass{llncs}
\usepackage{caption,url,alltt}
\usepackage[bookmarks=true,hyperfigures=true,colorlinks=true,linkcolor=blue,citecolor=blue,backref=page,pagebackref=false,pdfpagemode=fullscreen,plainpages=false,pdfpagelabels]{hyperref}
\usepackage[all]{hypcap}

\def\BigLaTeX{{\rm L\kern-.36em\raise.3ex\hbox{\smaller\smaller A}\kern-.15em
    T\kern-.1667em\lower.7ex\hbox{E}\kern-.125emX}}
\def\BoldLaTeX{{\bf L\kern-.36em\raise.3ex\hbox{\smaller\smaller\bf A}\kern-.15em
    T\kern-.1667em\lower.7ex\hbox{E}\kern-.125emX}}
\def\BibTeX{{\rm B\kern-.05em{\sc i\kern-.025em b}\kern-.08em
    T\kern-.1667em\lower.7ex\hbox{E}\kern-.125emX}}

\newcommand{\hozline}{{\noindent\rule{\textwidth}{0.4mm}}}

\newlength{\hsbw}
\newenvironment{session}{\begin{flushleft}
 \setlength{\hsbw}{\linewidth}
 \addtolength{\hsbw}{-\arrayrulewidth}
 \addtolength{\hsbw}{-\tabcolsep}
 \begin{tabular}{@{}|c@{}|@{}}\hline 
 \begin{minipage}[b]{\hsbw}
 \begingroup\sessionsize\vspace*{1.2ex}\begin{alltt}}{\end{alltt}\endgroup\end{minipage}\\ \hline
 \end{tabular}
 \end{flushleft}}
\def\extrawidth{0.5in}

\newenvironment{sessionlab}[1]{\begin{flushleft}
 \setlength{\hsbw}{\linewidth}
 \addtolength{\hsbw}{-\arrayrulewidth}
 \addtolength{\hsbw}{-\tabcolsep}
 \begin{tabular}{@{}|c@{}|@{}}\hline 
 \begin{minipage}[b]{\hsbw}
 \vspace*{-.5pt}
 \begin{flushright}
 \rule{0.01in}{.15in}\rule{0.9in}{0.01in}\hspace{-0.95in}
 \raisebox{0.04in}{\makebox[0.9in][c]{\footnotesize #1}}
 \end{flushright}
 \vspace*{-.47in}
 \begingroup\small\vspace*{1.0ex}\begin{alltt}}{\end{alltt}\endgroup\end{minipage}\\ \hline 
 \end{tabular}
 \end{flushleft}}
\newcounter{sessioncount}
\newenvironment{session*}{\begin{flushleft}
 \refstepcounter{sessioncount}
 \setlength{\hsbw}{\linewidth}
 \addtolength{\hsbw}{-\arrayrulewidth}
 \addtolength{\hsbw}{-\tabcolsep}
 \begin{tabular}{@{}|c@{}|@{}}\hline 
 \begin{minipage}[b]{\hsbw}
 \vspace*{-.5pt}
 \begin{flushright}
 \rule{0.01in}{.15in}\rule{0.3in}{0.01in}\hspace{-0.35in}
 \raisebox{0.04in}{\makebox[0.3in][c]{\footnotesize \thesessioncount}}
 \end{flushright}
 \vspace*{-.57in}
 \begingroup\small\vspace*{1.0ex}\begin{alltt}}{\end{alltt}\endgroup\end{minipage}\\ \hline 
 \end{tabular}
 \end{flushleft}}
\def\sessionsize{\small}
\def\smallsessionsize{\small}

\newcommand{\exmemo}[1]{}

\newcommand{\comment}[1]{}

\newcommand{\exfootnote}[1]{}
\newlength{\sblen}
\newlength{\overhang}

\def\SetFigFont#1#2#3{\rm}

\newcommand{\excite}[1]{}

\date{}
\makeatletter
\renewenvironment{verse}
               {\let\\\@centercr
                \list{}{\itemsep      \z@
                        \itemindent   -1.5em
                        \listparindent\itemindent
                        \advance\leftmargin 1.5em}
                \item\relax}
               {\endlist}
\makeatother
\usepackage[sectionbib,sort]{natbib}

\title{A Mechanically Assisted Examination\\ of Vacuity and Question  Begging\\ in
Anselm's Ontological Argument\thanks{This research was partially
    supported by SRI International.} 
}
\author{John Rushby}
\institute{Computer Science Laboratory\\ 
           SRI International\\ 
           333 Ravenswood Avenue\\ 
           Menlo Park, CA 94025 USA}
\begin{document}
\maketitle

\begin{abstract}

I use mechanized verification to examine several first- and
higher-order formalizations of Anselm's Ontological Argument against
the charge of begging the question.  I propose three different but
related criteria for a premise to beg the question in fully formal
proofs and find that one or another applies to all the formalizations
examined.  I also show that all these formalizations entail variants
that are vacuous, in the sense that they apply no interpretation to
``than which there is no greater'' and are therefore vulnerable to
Gaunilo's refutation.\\[-2.5ex]

My purpose is to demonstrate that mechanized verification provides an
effective and reliable technique to perform these analyses; readers
may decide whether the forms of question begging and vacuity so
identified affect their interest in the Argument or its various
formalizations.

\end{abstract}

\setcounter{section}{-1}
\section{Preamble}

This paper originally appears in ``Beyond Faith and Rationality:
Essays on Logic, Religion and Philosophy,'' edited by Ricardo
Silvestre, Paul Gocke, Jean-Yves Beziau and Purushottama Bilimoria,
published in Sophia Studies in Cross-cultural Philosophy of Traditions
and Cultures, vol 34, Springer, Sept.\ 2020, and it extends one
published in IfCoLog 5(7), 2018.

Recently, \cite{Oppenheimer&Zalta21}  published a paper
that, among other topics, criticizes my
formulation of ``begging the question'' and its application to the
Ontological Argument, so in this update to my paper I give more
intuitive explanations for my choices and conclusions and hope that
readers will find my case persuasive.

I have preserved the original content of the paper and
added the new material in sections marked as Addenda.

\section{Introduction}

I assume readers have some familiarity with St. Anselm's 11'th Century
Ontological Argument for the existence of God
\cite{Anselm:Proslogion}; a simplified translation from the original
Latin of Anselm's \emph{Proslogion} is given in Figure \ref{thearg},
with some alternative readings in square parentheses.  This version of
the argument appears in Chapter II of the Proslogion; another version
appears in Chapter III and speaks of the \emph{necessary} existence of
God.  Many authors have examined the Argument, in both its forms; in
recent years, most begin by rendering it in modern logic, employing
varying degrees of formality.  The Proslogion II argument is
traditionally rendered in first-order logic while propositional modal
logic is used for that of Proslogion III.  More recently, higher-order
logic and quantified modal logic have been applied to the argument of
Proslogion II\@.  My focus here is the Proslogion II argument,
represented completely formally in first- or higher-order logic, and
explored with the aid of a mechanized verification system.  Elsewhere,
I use a verification system to examine renditions of the argument in
modal logic \citep{Rushby:modalont19}, and also the argument of
Proslogion III \citep{Rushby21:ijpr}.

Verification systems are tools from computer science that are
generally used for exploration and verification of software or
hardware designs and algorithms; they comprise a specification
language, which is essentially a rich (usually higher-order) logic,
and a collection of powerful deductive engines (e.g., satisfiability
solvers for combinations of theories, model checkers, and automated
and interactive theorem provers).  I have previously explored
renditions of the Argument due to \cite{Oppenheimer&Zalta91} and
\cite{Eder&Ramharter15} using the PVS verification system
\citep{Rushby:ontological13,Rushby:ER-OntArg16}, and those provide the
basis for the work reported here.  \cite{Benzmueller&Paleo:ECAI14}
have likewise explored modal arguments due to G\"{o}del and Scott
using the Isabelle and Coq verification systems

\begin{figure}[t]
\hozline
\begin{verse}
1.  We can conceive of [something/that] than which there is no
greater\\[1ex]

2.  If that thing does not exist in reality, then we can conceive of a
greater\newline thing---namely, something [just like it] that does exist
in reality\\[1ex]

3.  Thus, either the greatest thing exists in reality or it is not\newline the
greatest thing\\[1ex]

4.  Therefore the greatest thing exists in reality\\[1ex]

[That's God]
\end{verse}
\hozline
\caption{\label{thearg}The Ontological Argument}
\end{figure}

Mechanized analysis confirms the conclusions of most earlier
commentators: the Argument is valid.  Attention therefore focuses on
the premises and their interpretation.  The premises are \emph{a
priori} (i.e., armchair speculation) and thus not suitable for
empirical confirmation or refutation: it is up to the individual
reader to accept or deny them.  We may note, however, that the
premises are consistent (i.e., they have a model), and this is among
the topics that I previously subjected to mechanized examination
\citep{Rushby:ontological13} (as a byproduct, this examination
demonstrates that the Argument does not compel a theological
interpretation: in the exhibited model, that ``than which there is no
greater'' is the number zero).

The Argument has been a topic of enduring fascination for nearly a
thousand years; this is surely due to its derivation of a bold
conclusion from unexceptionable premises, which naturally engenders a
sense of disquiet: \citet[page 472]{Russell:HistoryWP} opined ``The
Argument does not, to a modern mind, seem very convincing, but it is
easier to feel that it must be fallacious than it is to find out
precisely where the fallacy lies.''  Many commentators have sought to
identify a fallacy in the Argument or its interpretation (e.g., Kant
famously denied it on the basis that ``existence is not a
predicate'').  One direction of attack is to claim that the Argument
``begs the question''\footnote{This phrase is widely misunderstood to
mean ``to invite the question.''  Its use in logic derives from
medieval translations of Aristotle, where the Latin form \emph{Petitio
Principii} is also employed.}; that is, it essentially assumes what it
sets out to prove \citep{Rowe76,Walton78}.  This is the primary charge
that I examine here.

Begging the question has traditionally been discussed in the context
of informal or semi-formal argumentation and dialectics
\citep{Barker76,Barker78,Sanford77,Walton91,Walton94,Walton06}, where
it is debated whether arguments that beg the question should be
considered fallacious, or valid but unpersuasive, or may even be
persuasive.  Here, we examine question begging in the context of fully
formal, mechanically checked proofs.  My purpose is to provide
techniques that can identify potential question begging in a
systematic and fairly unequivocal manner.  I do not condemn the forms
of question begging that are identified; rather, my goal is to
highlight them so that readers can make up their own minds and can
also use these techniques to find other cases.

A secondary charge that I will consider is one of \emph{vacuity}: most
of the formalized arguments examined here entail variants that apply
no interpretation to ``than which there is no greater.''  We are
therefore free to apply any interpretation and in this way can
reproduce the ``lost island'' parody that Gaunilo
\citeyearpar{Gaunilo} used to claim refutation of the original
argument.

The chapter is structured as follows.  In the next section, I
introduce a strict definition of ``begging the question'' and show
that a rendition of the Argument due to 
\cite{Oppenheimer&Zalta91} is vulnerable to this charge.  Oppenheimer
and Zalta use a definite description (i.e., they speak of
``\emph{that} than which there is no greater'') and require an
additional assumption to ensure this is well-defined.  \citet[Section
2.3]{Eder&Ramharter15} argue that Anselm did not intend this
interpretation (i.e., requires only ``\emph{something} than which
there is no greater'') and therefore dispense with the additional
assumption of Oppenheimer and Zalta.  In Section \ref{weak}, I show
that this version of the argument does not beg the question under the
strict definition, but that it does so under a plausible weakening.  I
then turn to the topic of vacuity and, in Section \ref{vacuity}, show
that this version of the argument has a variant that applies no
interpretation to ``than which there is no greater'' and is thus
vulnerable to Gaunilo's refutation; I argue that the original
formulation shares this defect.  In Section \ref{indirect}, I consider
an alternative premise due to Eder and Ramharter and show that this
does not beg the question under either of the previous
interpretations, but I argue that it is at least as questionable as
the premise that it replaces because it so perfectly discharges the
main step of the proof that it seems reverse-engineered.  I suggest a
third interpretation for ``begging the question'' that matches this
case.  In Section \ref{ho}, I consider the higher-order treatment of
\citet[Section 3.3]{Eder&Ramharter15} and a variant derived from
\cite{Campbell18:book}; these formalized proofs are more complicated
than those of the first-order treatments but I show how the third
interpretation for ``begging the question'' applies to them.  I also
show that all these versions have vacuous variants.  In Section
\ref{compare}, I compare my interpretations to existing, mainly
informal, accounts of what it means to ``beg the question.''  Finally,
in Section \ref{conc}, I summarize and show that all the
formalizations examined can be generated as elaborations of a
manifestly vacuous and circular starting argument.

\section{Begging the Question: Strict Case}
\label{direct}

``Begging the question'' is a form of circular reasoning in which we
assume what we wish to prove.  It is generally discussed in the
context of informal argumentation where the premises and conclusion
are expressed in natural language.  In such cases, the
question-begging premise may state the same idea as the conclusion,
but in different terms, or it may contain superfluous or even false
information, and there is much literature on how to diagnose and
interpret such cases
\citep{Barker76,Barker78,Sanford77,Walton94,Walton91,Walton06}.  That
is not my focus.  I am interested in formal, deductive arguments, and
in criteria for begging the question that are themselves formal.  Now,
deductive proofs do not generate new knowledge---the conclusion is
always implicit in the premises---but they can generate surprise and
they can persuade; I propose that criteria for question begging should
focus on the extent to which either the conclusion or its proof are
``so directly'' represented in the premises as to vitiate the hope of
surprise or persuasion.

\vspace*{-2ex}
\subsubsection*{Addendum} to the published paper.

Before proceeding to our formal examination, let us consider the
informal presentation of the Argument in Figure \ref{thearg} and ask
whether it begs the question in the sense described above.  We can
abbreviate the argument by combining the lines numbered 2 and 3 as
follows.\footnote{This combination of lines 2 and 3 elides one of the
more interesting aspects of Anselm's argument: namely that the
alternative thing to consider if the first does not exist in reality
is something \emph{just like it} that does.  The first-order
formalizations with which we begin our examination do not (cannot)
express this construction and so I consider the combination is
reasonable at this stage.  The higher-order formalizations of Section
\ref{ho} hew closer to Anselm's construction, but modal logic is the
preferred vehicle for more faithful formalizations
\citep{Rushby:modalont19}.}

\hozline
\begin{verse}
1.  We can conceive of something than which there is no
greater (call that ``the greatest'')

2+3. If a thing does not exist in reality, then it is not the greatest

4.  Therefore the greatest thing exists in reality
\end{verse}
\vspace*{-1.5ex}
\hozline

\noindent And now we can replace the new line 2+3 by its
contrapositive as follows.\\[-1ex]
\hozline
\begin{verse}
1.  We can conceive of something than which there is no
greater (call that ``the greatest'')

2+3$'$.  If a thing is the greatest, then it exists in reality

4.  Therefore the greatest thing exists in reality
\end{verse}
\vspace*{-1.5ex}
\hozline

Now this form of the argument is surely trivial and of little
interest: Premise 2+3$'$ states that the other premise implies the
conclusion.  Thus, ``the conclusion is `so directly' represented in
the premises as to vitiate the hope of surprise or persuasion'' and
therefore this argument begs the question by that criterion.

So how does this apply to the original form of the argument, before we
took the contrapositive?  I suggest there are two ways of looking at
this question.  One holds that the original argument is just an
obfuscation of the trivial one and that once the obfuscation is
revealed, the original likewise loses all hope of surprise or
persuasion and therefore is also considered question begging.  The
other point of view is that although the contraposed premise 2+3$'$
makes the argument logically trivial, the soundness of the argument
depends on whether we believe that premise, which I at least find
dubious.  Premise 2+3 of the original argument, on the other hand,
does suggest why we should believe it and, therefore, preserves some
element of persuasion and can be excused of begging the question.

My focus in this paper is on systematic ways to define and detect
question begging of this kind in formalized arguments.  It is up to
the reader to decide if this is a useful capability and whether the
properties so revealed really do amount to question begging and
whether they should cause concern.  At the very least, I hope these
investigations validate Russell's observation that ``the Argument does
not \ldots seem very convincing'' by revealing ``how the trick is
done.''  \textbf{End of addendum}\\[-1ex]

The most elementary instance of begging the question is surely when
one, or a collection, of the premises is equivalent to the conclusion.
But if some of the premises are equivalent to the conclusion, what are
the other premises for?  Certainly we must need all the premises to
deduce the conclusion (else we can eliminate some of them); thus we
surely need all the premises before we can establish that some of them
are equivalent to the conclusion.  Hence, criteria for begging the
question should apply \emph{after} we have accepted the other
premises.  Thus, if $C$ is our conclusion, $Q$ our ``questionable''
premise (which may be a conjunction of simpler premises) and $P$ our
other premises, then $Q$ begs the question in this elementary or
\emph{strict} sense if $C$ is equivalent to $Q$, assuming $P$\/: i.e.,
$P \vdash C \equiv Q$ (we use $\vdash$ for ``proves,'' $\equiv$ for
equivalence and, later, $\supset$ for material implication).  Of
course, this means we can prove $C$ using $Q$: $P, Q \vdash C$, and we
can also do the reverse: $P, C \vdash Q$.

\begin{figure}[!ht]
\begin{session}\small
OandZ: THEORY
BEGIN

  beings: TYPE
  x, y: VAR beings

  >: (trichotomous?[beings])

  God?(x): bool = NOT EXISTS y: y > x

  re?(x): bool

  ExUnd: AXIOM EXISTS x: God?(x)

  Greater1: AXIOM FORALL x: (NOT re?(x) => EXISTS y: y > x)

  God_re: THEOREM re?(the(God?))


  Greater1_circ: THEOREM God_re IMPLIES Greater1

END OandZ
\end{session}
\vspace*{-1ex}
\caption{\label{oandz}Oppenheimer and Zalta's Treatment, in PVS}
\end{figure}

Figure \ref{oandz} presents Oppenheimer and Zalta's
\citeyearpar{Oppenheimer&Zalta91} treatment of the Ontological
Argument formalized in PVS, using the notation of 
\citet[Section 3.2]{Eder&Ramharter15}.  I will not describe this formal
specification in detail, since it is done at tutorial level elsewhere
\citep{Rushby:ontological13}, but I will explain the basic language and
ideas.  Briefly, the specification language of PVS is a strongly typed
higher-order logic with predicate subtypes.  This example uses only
first order but does make essential use of predicate subtypes and the
proof obligations that they can incur \citep{Rushby98:TSE}.  The
uninterpreted type \texttt{beings} is used for those things that are
``in the understanding.''  Note that a question mark at the end of an
identifier is merely a convention to indicate predicates (which in PVS
are simply functions with return type \texttt{bool}).  A predicate in
parentheses denotes the corresponding predicate subtype, so that
\texttt{>} is an uninterpreted relation on \texttt{beings} that
satisfies the predicate \texttt{trichotomous?}, which is part of the
``Prelude'' of standard theories built in to PVS.\footnote{Trichotomy
is the condition \texttt{FORALL x, y: x > y OR y > x OR x = y}.}  PVS
generates a proof obligation (not shown here) called a Typecheck
Correctness Condition, or TCC, to ensure such a relation exists, which
we discharge by exhibiting the everywhere true relation.

The predicate \texttt{God?}\ recognizes those beings ``than which
there is no greater''; the axiom \texttt{ExUnd} asserts the existence
of at least one such being; \texttt{the(God?)}  is a definite
description that identifies this being.  PVS generates a TCC (not
shown here) to ensure this being exists and is unique (this is
required by the predicate subtype used in the definition of
\texttt{the}, which is part of the PVS Prelude), and \texttt{ExUnd}
and the trichotomy of \texttt{>} are used to discharge this
obligation.  The uninterpreted predicate \texttt{re?}\ identifies
those beings that exist ``in reality'' and the axiom \texttt{Greater1}
asserts that if a being does not exist in reality, then there is a
greater being.  Note that the string \texttt{IMPLIES} and the symbol
\texttt{=>} are entirely equivalent in PVS (also the string
\texttt{AND} and symbol \texttt{\&}); we use whichever seems
most readable in its context.

The theorem \texttt{God\_re} asserts that the being identified by the
definite description \texttt{the(God?)} exists in reality.  The PVS
proof of this theorem is accomplished by the following commands.
\begin{sessionlab}{PVS Proof}
(typepred "the(God?)")  (use "Greater1") (grind)
\end{sessionlab}
These commands invoke the type associated with \texttt{the(God?)}
(namely that it satisfies the predicate \texttt{God?}), the premise
\texttt{Greater1}, and then apply the standard automated proof
strategy of PVS, called \texttt{grind}.  Almost all the proofs
mentioned subsequently are similarly straightforward and we do not
reproduce them in detail.

As first noted by  \cite{Garbacz12}, the premise
\texttt{Greater1} begs the question under the other assumptions of the
formalization.  We state the key implication as
\texttt{Greater1\_circ} (PVS Version 7 allows formula names to be used
in expressions as shorthands for the formulas themselves) and prove it
as follows.
\begin{sessionlab}{PVS proof}
 (expand "God_re")
 (expand "Greater1")
 (typepred "the(God?)")
 (grind :polarity? t)
 (inst 1 "x!1")
 (typepred ">")
 (grind)
\end{sessionlab}
The first two steps expand the formula names to the formulas
they represent, the \texttt{typepred} steps introduce the predicate
subtypes associated with their arguments (namely, that
\texttt{the(God?)}  satisfies \texttt{God?} and that \texttt{>} is
trichotomous) and the other steps perform quantifier reasoning and
routine deductions.

Given that we have proved \texttt{God\_re} from \texttt{Greater1} and
vice-versa, we can easily prove they are equivalent.  Thus, in the
definition of ``begging the question'' given earlier, $C$ here is
\texttt{God\_re}, $Q$ is \texttt{Greater1} and $P$ is the rest of the
formalization (i.e., \texttt{ExUnd}, the definition of \texttt{God?}
and the definite description \texttt{the(God?)}, and the predicate
subtype \texttt{trichotomous?}\ asserted for \texttt{>}).

Notice that we can also prove the premise \texttt{ExUnd} from the
conclusion \texttt{God\_re}, so it looks as if \texttt{ExUnd} begs the
question, too.  However, \texttt{ExUnd} is used to discharge the TCC
that ensures the definite description operator \texttt{the} is used
appropriately.  Hence, \texttt{ExUnd} is strictly prior to
\texttt{God\_re} (because the specification is not accepted until its
TCCs are discharged), thereby breaking the circularity.  Hence,
\texttt{ExUnd} cannot be considered to beg the question in this case.

\subsubsection*{Addendum} Recently, \cite{Oppenheimer&Zalta21}
attempted to rebut my claim that \texttt{Greater1} is question
begging.  Most of their points were anticipated in Section
\ref{compare} of my paper as published (and which is extended here),
but I was remiss in not connecting the formal notion of strict
begging, as established for \texttt{Greater1}, to an intuitive
explanation why this premise should be considered to beg the question.

To see this, we perform a similar transformation as that applied to
the informal rendition of the Argument in the addendum at the
beginning of this section.  That is, we observe that \texttt{Greater1}
could be replaced by its contrapositive
\begin{session}
  Greater1cp: LEMMA FORALL x: (NOT EXISTS y: y > x) => re?(x)
\end{session}
and observe further that the left side of the implication is just
\texttt{God?(x)} and so the formula can be rewritten as follows.
\begin{session}
  Greater1cp_alt: LEMMA FORALL x: God?(x) => re?(x)
\end{session}
My informal criterion for question begging, stated at the start of
this section, concerns ``the extent to which either the conclusion or
its proof are `so directly' represented in the premises as to vitiate
the hope of surprise or insight.''  All will surely agree that the
conclusion \texttt{God\_re} is ``so directly'' represented in the
premise \texttt{Greater1cp\_alt} (since it says that the other
premise, \texttt{ExUnd}, directly entails the conclusion)
that there can be no surprise or insight, and so this premise begs the
question.  But \texttt{Greater1} is just an obfuscated version of
\texttt{Greater1cp\_alt} so it, too, surely begs the question.

As we will discuss in Section \ref{compare},
Eder and Ramharter \cite[Section 1.2(5)]{Eder&Ramharter15} observe
that the conclusion to a deductively valid argument must be implicit
or ``contained'' in the premises (otherwise, the reasoning would not
be deductive), but an argument can only be persuasive or interesting
``if it is possible to accept the premises without already recognizing
that the conclusion follows from them.  Thus, the desired conclusion
has to be `hidden' in the premises.''  In a footnote, they aver
``Sometimes, proofs of the existence of God are accused of being
question-begging, but this critique is untenable.  It is odd to ask
for a deductive argument whose conclusion is not contained in the
premises.  Logic cannot pull a rabbit out of the hat.''

Eder and Ramharter are, of course, correct that the conclusion must be
``contained'' in the premises, but they are also correct that it
should be ``hidden,'' and so I dispute their claim that accusations of
question begging are untenable for the Ontological Argument.  I
suggest that tests for question begging should expose the ``hiding
place'' of the conclusion among the premises: if this is revealed as
inadequate or contrived, then our interest in the argument, and its
persuasiveness, are diminished.  ``Hiding'' generally takes the form
of obfuscation, and I suggest this may be considered flimsy when it
uses only propositional rearrangement or definition
expansion/folding, but no quantifier reasoning.

Recall the notation where $C$ is our conclusion, $Q$ our
``questionable'' premise, $P$ our other premises, and that $P, Q
\vdash C$.  Then the deduction theorem gives $P \supset (Q \supset C)$
although $Q$ may be expressed in a form that ``hides'' or obfuscates
this relationship.  The formal criterion of strict begging strengthens
the right hand side of the first implication to $Q \equiv C$ and this
reduces the opportunity for obfuscation (e.g., $Q$ cannot be stronger
than required) and exposes those question begging premises that are
not well hidden.  Here, strict begging has identified that the
conclusion is ``hiding'' in \texttt{Greater1}, and
\texttt{Greater1cp\_alt} reveals that its cover is rather flimsy.
\textbf{End of addendum}

\section{Begging the Question: Weaker Case}
\label{weak}

\begin{figure}[!t]
\begin{session}\small
EandR1: THEORY
BEGIN

  beings: TYPE
  x, y: VAR beings

  >(x, y): bool

  God?(x): bool = NOT EXISTS y: y > x

  re?(x): bool

  ExUnd: AXIOM EXISTS x: God?(x)

  Greater1: AXIOM FORALL x: (NOT re?(x) => EXISTS y: y > x)

  God_re_alt: THEOREM EXISTS x: God?(x) AND re?(x)


  Greater1_circ1: THEOREM trichotomous?(>)
     IMPLIES God_re_alt => Greater1

  Greater1_circ2: THEOREM (FORALL x, y: God?(x) => x>y or x=y)
     IMPLIES God_re_alt => Greater1

END EandR1
\end{session}\vspace*{-1ex}
\caption{\label{eandr1}Eder and Ramharter's First Order Treatment, in PVS}
\end{figure}

\citet[Section 2.3]{Eder&Ramharter15} claim that
Anselm's Proslogion does not employ a definite description and that a
correct reading is ``\emph{something} than which there is no
greater.''  A suitable modification to the previous PVS theory is
shown in Figure \ref{eandr1}; the differences are that \texttt{>} is
now an unconstrained relation on \texttt{beings}, and the conclusion
is restated as the theorem \texttt{God\_re\_alt}.  As before, this
theorem is easily proved from the premises \texttt{ExUnd} and
\texttt{Greater1} and the definition of \texttt{God?}.  And also as
before, the premise \texttt{ExUnd} can be proved from the conclusion
\texttt{God\_re\_alt}, so this premise strictly begs the question
(unlike the version of Figure \ref{oandz}, there are no TCCs here to
break the circularity).  However, \texttt{Greater1} is no longer
strictly begging because it cannot be proved from the conclusion
\texttt{God\_re\_alt}.

We can observe, however, that this specification of the Argument is
very austere and imposes no constraints on the relation \texttt{>}; in
particular, it could be an entirely empty relation.  We demonstrate
this in the theory interpretation \texttt{EandR1interp} shown in
Figure \ref{eandr1-model}, where beings are interpreted as natural
numbers, all beings exist in reality, and none are \texttt{>} than any
other; thus, any natural number satisfies \texttt{God?}.  PVS
generates proof obligations (not shown here) to ensure the axioms of
the theory \texttt{EandR1} are theorems under this interpretation, and
these are trivially true.

\begin{figure}[!t]
\begin{session}\small
EandR1interp: THEORY
BEGIN

IMPORTING EandR1\{\{ 
  beings := nat,
  re? := LAMBDA (x: nat): TRUE,
  > := LAMBDA (x, y: nat): FALSE
\}\} AS model

END EandR1interp
\end{session}\vspace*{-1.6ex}
\caption{\label{eandr1-model}The Empty Model for Eder and Ramharter's
First Version}
\end{figure}

Such a model seems contrary to the intent of the Argument: surely it
is not intended that something than which there is no greater is so
because nothing is greater than anything else.  So we should
require some minimal constraint on \texttt{>} to eliminate such
impoverished models.  A plausible constraint is that \texttt{>} be
trichotomous; if we add this condition, as in
\texttt{Greater1\_circ1}, then the premise \texttt{Greater1} can
again be proved from the conclusion \texttt{God\_re\_alt}.  A weaker
condition is to require only that beings satisfying the \texttt{God?}\
predicate should stand in the \texttt{>} relation to others; this is
stated in \texttt{Greater1\_circ2} and is also sufficient to
prove \texttt{Greater1} from \texttt{God\_re\_alt}.

In terms of the abstract formulation given at the beginning of Section
\ref{direct}, what we have here is that the conclusion $C$ can be
proved using the questionable premise $Q$: $P, Q \vdash C$, but not
\emph{vice versa}.  However, if we augment the other premises $P$ by
adding some $P_2$, then we can indeed prove $Q$: $P, P_2, C \vdash Q$,
and also the equivalence of $C$ and $Q$: $P, P_2 \vdash C \equiv Q$.  Thus,
$Q$ does not beg the question $C$ under the original premises $P$ but
does do so under the augmented premises $P, P_2$.  In this case, we
will say that $Q$ \emph{weakly begs} the question, where $P_2$
determines the ``degree'' of weakness.

In this example, the question begging premise fails our definition of
strict begging because it is used in an impoverished theory, and weak
begging compensates for that.  Another way a premise $Q$ can escape
strict begging is by being stronger than necessary and one way to
compensate for that is to strengthen the conclusion by conjoining some
$S$ so that $P, (C \wedge S) \vdash Q$ and $P, Q \vdash (C \wedge S)$.
However, it may be difficult to satisfy both of these simultaneously
and the first is equivalent to weak begging with $P_2 = S$; hence, we
prefer the original, more versatile, notion of weak begging.

The rationale for introducing weak begging is that it exposes strict
begging that is otherwise masked by an impoverished theory or a strong
premise.  But with enough deductive power we can always construct a
$P_2$ and thereby claim weak begging; the question is whether this
additional premise is plausible and innocuous in the intended
interpretation, and this is a matter for human judgment.

\subsubsection*{Addendum}
As with the version of Oppenheimer and Zalta, our analysis identifies
\texttt{Greater1} as the ``hiding place'' for question begging, and
folding the definition of
\texttt{God?(x)} (see \texttt{Greater1\_triv} below) and then
taking the contrapositive exposes the obfuscation
employed.

I should also have observed that in this example \texttt{ExUnd}
satisfies the criterion for strict begging, but it is obviously
unreasonable to accuse it of begging the question because it is needed
to supply the witness for \texttt{x} in the conclusion.  My formal
definitions for question begging are not unequivocal: they identify
candidates that might beg the question, but human judgment is required
to decide the matter.  \textbf{End of addendum}

\section{Trivializing the Argument, and Gaunilo's Refutation}
\label{vacuity}

Notice that the right side of the implication in \texttt{Greater1} of
Figure \ref{eandr1} is equivalent to \texttt{NOT God?(x)}, so that
\texttt{Greater1} can be rewritten as \texttt{Greater1\_triv}, as
shown below.
\begin{sessionlab}{PVS fragment}
  Greater1_triv: LEMMA
    FORALL x: (NOT re?(x) => NOT God?(x))
\end{sessionlab}
But now we can prove the theorem
without opening the definition of \texttt{God?}.  We do this as follows.
\begin{sessionlab}{PVS proof}
(lemma "ExUnd") (lemma "Greater1_triv")
(grind :exclude "God?")
\end{sessionlab}
Here, we install the premises \texttt{ExUnd} and
\texttt{Greater1\_triv}, and then invoke the general purpose strategy
\texttt{grind}, instructing it not to open the definition of
\texttt{God?}.  Of course, we hardly need mechanized theorem proving
to verify this: \texttt{ExUnd} says there is some being satisfying
\texttt{God?}, the contrapositive of \texttt{Greater1\_triv} says such
a being satisfies \texttt{re?}, and we are done.  

This is not only a trivial argument, but it does not depend on the
meaning attached to ``God'' and so we could replace it by any other
term and interpretation.  In particular, we could substitute ``the
most perfect island'' and thereby reproduce the ``lost island'' parody
by Gaunilo, a contemporary of Anselm, who used the form of Anselm's
argument to establish the (absurd) existence of that most perfect
island (Gaunilo \citeyear{Gaunilo}). Anselm and other authors defend
the informal Proslogion II argument against Gaunilo's
parody,\footnote{One approach asserts that a contingent object, such
as an island, can always be improved and thus could never be one than
which there is no greater.}  but the formalization of Figure
\ref{eandr1} with \texttt{Greater1\_triv} is indefensible, since it is
true for all interpretations.  We will say that such an argument is
\emph{vacuous}.  For later reference, we show this vacuous form of the
argument in Figure \ref{vac1}; \texttt{Greater1\_vac} is the
contrapositive of \texttt{Greater1\_triv}.

\begin{figure}[!t]
\begin{session}\small
Vacuous: THEORY
BEGIN

  beings: TYPE
  x, y: VAR beings

  God?(x): bool

  re?(x): bool

  ExUnd: AXIOM EXISTS x: God?(x)

  Greater1_vac: AXIOM FORALL x: God?(x) => re?(x) 

  God_re_alt: THEOREM EXISTS x: God?(x) AND re?(x)

END Vacuous
\end{session}\vspace*{-1ex}
\caption{\label{vac1}A Vacuous Version of the Argument in PVS}
\end{figure}

We should now ask whether the vacuity of Figure \ref{vac1} and of
Figure \ref{eandr1} with \texttt{Greater1\_triv} also applies to the
original specification with \texttt{Greater1}.  My opinion is that it
does, because we can systematically transform Figure \ref{vac1} into
Figure \ref{eandr1}: we simply apply an interpretation to
\texttt{God?} and open up its appearance in \texttt{Greater1\_vac} to
reveal that interpretation, then take the contrapositive and thereby
obtain \texttt{Greater1}.  It is irrelevant what the interpretation
is, so the symbol \texttt{>} and its reading as ``greater'' are
entirely specious, and we cannot attach any belief to
\texttt{Greater1} once we see this derivation.  I will return to this
topic in the Conclusion, Section \ref{conc}.

It is worth noting that Oppenheimer and Zalta's specification of
Figure \ref{oandz} cannot be reduced to a similarly vacuous form
because it becomes impossible to discharge the TCC proof obligation
that is required to ensure that the definite description is unique.

\section{Indirectly Begging the Question}
\label{indirect}

\citet[Section 3.2]{Eder&Ramharter15} consider \texttt{Greater1} an
unsatisfactory premise because it does not express ``conceptions
presupposed by the author'' (i.e., Anselm) and says nothing about what
it means to be \emph{greater} other than the contrived connection to
\emph{exists in reality}.  They propose an alternative premise
\texttt{Greater2}, which is shown in Figure \ref{eandr2}.  This theory
is the same as that of Figure \ref{eandr1}, except that
\texttt{Greater2} is substituted for \texttt{Greater1}, and a new
premise \texttt{Ex\_re} is added.

\begin{figure}[!ht]
\begin{session}\small
EandR2: THEORY
BEGIN

  beings: TYPE
  x, y: VAR beings

  >(x, y): bool

  God?(x): bool = NOT EXISTS y: y > x

  re?(x): bool

  ExUnd: AXIOM EXISTS x: God?(x)

  Ex_re: AXIOM EXISTS x: re?(x)

  Greater2: AXIOM FORALL x, y: (re?(x) AND NOT re?(y) => x > y)

  God_re_alt: THEOREM EXISTS x: God?(x) AND re?(x)

END EandR2
\end{session}
\caption{\label{eandr2}Eder and Ramharter's Adjusted First Order Treatment, in PVS}
\end{figure}

Before we proceed to examine question begging in this version, we can
note that the right side of the implication in \texttt{Greater2}
entails \texttt{NOT God?(y)}, so the premise entails the following
variant.\footnote{The published paper incorrectly states these
premises are equivalent.}
\begin{sessionlab}{PVS fragment}
  Greater2_triv: LEMMA
    FORALL x, y: (re?(x) AND NOT re?(y) => NOT God?(y))
\end{sessionlab}
This variant premise, plus \texttt{ExUnd} and \texttt{Ex\_re} can be
used to prove the conclusion \texttt{God\_re\_alt} without opening the
definition of \texttt{God?}.  Thus, the theory with
\texttt{Greater2\_triv} in place of \texttt{Greater2} is vacuous and,
by the same reasoning as in Section \ref{vacuity}, I argue that the
original Figure \ref{eandr2} is too.

Next, we return to the original premises with \texttt{ExUnd},
\texttt{Ex\_re} and \texttt{Greater2}, and note that these also prove
the conclusion \texttt{God\_re\_alt} and that \texttt{ExUnd} and
\texttt{Ex\_re} strictly beg the question.  These three premises also
entail \texttt{Greater1} of Figure \ref{eandr1}, so there is
circumstantial evidence that \texttt{Greater2} is question begging.
However, it is not possible to prove \texttt{Greater2} from
\texttt{God\_re\_alt} and the other premises, nor have I found a
plausible augmentation to the premises that enables this.  Thus, it
seems that \texttt{Greater2} does not beg the question under our
current definitions, neither strictly nor weakly.

However, when constructing a mechanically checked proof of
\texttt{God\_re\_alt} using \texttt{Greater2} I was struck how neatly
the premise exactly fits the requirement of the interactive proof at its
penultimate step.  To see this, observe the PVS sequent shown below;
we arrive at this point following a few straightforward steps in the
proof of \texttt{God\_re\_alt}.  First, we introduce the premises
\texttt{ExUnd} and \texttt{Ex\_re}, expand the definition of
\texttt{God?}, and perform a couple of routine steps of Skolemization,
instantiation, and propositional simplification.

\begin{sessionlab}{PVS Sequent A}
God_re_alt :  

[-1]  re?(x!1)
  |-------
\{1\}   x!1 > x!2
[2]   re?(x!2)

\end{sessionlab}
PVS represents its current proof state as the leaves of a tree of
sequents (here there is just one leaf); each sequent has a collection
of numbered formulas above and below the \texttt{|-----} turnstile
line; the interpretation is that the conjunction of formulas above the
line should entail the disjunction of those below.  Bracketed numbers
on the left are used to identify the lines, and braces (as opposed to
brackets) indicate this line is new or changed since the previous
proof step.  Terms such as \texttt{x!1} are Skolem constants.  PVS
eliminates top level negations by moving their formulas to the other
side of the turnstile.  Thus the sequent above is equivalent to the
following.
\begin{sessionlab}{Variant Sequent}
God_re_alt :  

[-1]  re?(x!1)
[2]   NOT re?(x!2)
  |-------
\{1\}   x!1 > x!2

\end{sessionlab}
We can read this as
\begin{session}
re?(x!1) AND NOT re?(x!2) IMPLIES x!1 > x!2
\end{session}
and then observe that \texttt{Greater2} is its universal generalization.

PVS has capabilities that help mechanize this calculation.  If we ask
PVS to generalize the Skolem constants in the original sequent A, it
gives us the formula
\begin{session}
FORALL (x_1, x_2: beings): re?(x_2) IMPLIES x_2 > x_1 OR re?(x_1)
\end{session}
Renaming the variables and rearranging, this is
\begin{session}
FORALL (x, y: beings): (re?(x) AND NOT re?(y)) IMPLIES x > y
\end{session}
which is identical to \texttt{Greater2}.   Thus, \texttt{Greater2}
corresponds \emph{precisely} to the formula required to discharge the
final step of the proof.

I will say that a premise \emph{indirectly} begs the question if it
supplies exactly what is required to discharge a key step in the
proof.  Unless they are redundant or superfluous, all the premises to
a proof will be essential to its success, so it may seem that any
premise can be considered to indirectly beg the question.
Furthermore, if we do enough deduction, we can often arrange things so
that the final premise to be installed exactly matches what is
required to finish the proof.  My intent is that the criterion for
indirect begging applies only when the premise in question perfectly
matches what is required to discharge a key (usually final) step of
the proof when the preceding steps have been entirely routine.  It is
up to the individual to decide what constitutes ``routine'' deduction;
I include Skolemization, propositional simplification, definition
expansion and rewriting, but draw the line at nonobvious quantifier
instantiation.  The current example does require quantifier
instantiation: a few steps prior to Sequent A above, the proof state
is represented by the following sequent.
\begin{sessionlab}{PVS Sequent}
God_re_alt :  

\{-1\}  God?(x!1)
\{-2\}  re?(x!2)
  |-------
[1]   EXISTS x: God?(x) AND re?(x)

\end{sessionlab}
The candidates for instantiating \texttt{x} are the Skolem constants
\texttt{x!1} or \texttt{x!2}.  The correct choice is \texttt{x!1} and
I would allow this selection, or even some experimentation with
different choices, within the ``obvious'' threshold, though others may
disagree.

I claim that the sequent constructed by the PVS prover following
routine deductions is a good representation of our epistemic state
after we have digested the other premises.  If the questionable
premise supplies exactly what is required to complete the proof from
that point (by generalizing the sequent), then it cannot be understood
independently and therefore satisfies the ``epistemic'' criterion for
question begging \citep{Walton94}, to be discussed in Section
\ref{compare}.  Furthermore, its construction appears
reverse-engineered and this eliminates any hope of surprise or
persuasion and thereby satisfies another characteristic of question
begging.

My description of indirect begging is very operational and might seem
tied to the particulars of the PVS prover, so we can seek a more
abstract definition.  After we have installed the other premises, the
PVS sequent is a representation of $P \supset C$.  The proof
engineering that reveals $Q$ indirectly to beg the question shows that
$Q$ is what is needed to make this a theorem, so $\vdash Q \supset (P
\supset C)$.  But more than this, it is \emph{exactly} what is needed,
so we could suppose $\vdash Q \equiv (P \supset C)$ and then take this
as a definition of indirect begging.\exfootnote{I am grateful to one
of the reviewers for this suggestion.}  Notice that this definition
implies strict begging, but not vice-versa.  However, a difficulty
with this definition is that the direction $\vdash (P \supset C)
\supset Q$ is generally stronger than can be proved.  The proof
engineering approach to indirect begging can be seen as an operational
way to interpret and approximate this definition: we use deduction to
simplify $P \supset C$ and then ask whether $Q$ is its universal
generalization.

In simple cases, the proof engineering approach is straightforward and
makes good use of proof automation, but it may be difficult to apply
in more complex proofs where a premise is employed as part of a longer
chain of deductions.  In the following section I show how careful
proof structuring can, without undue contrivance, isolate the
application of a premise and expose its question begging character.

\subsubsection*{Addendum} Again, I failed to provide an intuitive
explanation why \texttt{Greater2} should be considered to beg the
question.  An indirect reason is that \texttt{Ex\_re?} and
\texttt{Greater2} together entail \texttt{Greater1}, which we have
already established to be question begging.  A more direct explanation
is to note that \texttt{Greater2} is stronger than required and we can
weaken it by existentially quantifying the \texttt{x} on the right
hand side of the implication; but then that right hand side is just
\texttt{NOT God?(y).}  We can then do some propositional rearrangement
of the formula to yield \texttt{Greater2\_circ} as shown below.
\begin{session}
  Greater2_circ: COROLLARY re?(x) AND God?(y) => re?(y)
\end{session}
This surely begs the question, for it says that the other two premises
(which supply \texttt{re?(x)} and \texttt{God?(y)}) directly imply the
conclusion.  The original \texttt{Greater2} is simply a strengthened
and obfuscated version of \texttt{Greater2\_circ} and inherits its
question begging character.

I am tempted to argue the general case: any formula that entails an
obviously question begging premise should itself be considered
question begging, but this goes too far.  Recall the notation where
$C$ is our conclusion, $Q$ our ``questionable'' premise and $P$ our
other premises, then an ``obviously question begging'' premise would
be one that states that the other premises directly entail the
conclusion: namely, $P \supset C$.  Hence, we would indict any $Q$ such
that $Q \supset (P \supset C)$; but we have $Q, P \vdash C$ and the
previous formula then follows by the deduction theorem and so we would
indict every $Q$.  I suggest we can rescue something from this line of
argument and indict just those $Q$ that entail the obviously question
begging formula using only propositional rearrangement and
folding/expansion of definitions.  This would leave \texttt{Greater2}
on the borderline, since its entailment of the obviously question
begging premise uses quantificational reasoning, but of a trivial kind
(no search is required: all the instantiations are forced).
\textbf{End of addendum}

\section{Indirect Begging in More Complex Proofs}
\label{ho}

In search of a more faithful reconstruction of Anselm's Argument,
\citet[Section 3.3]{Eder&Ramharter15} observe that Anselm attributes
properties to beings and that some of these (notably \emph{exists in
reality}) contribute to evaluation of the \emph{greater} relation.
They formalize this by hypothesizing some class \texttt{P} of
``\emph{greater}-making'' properties on beings and then define one
being to be greater than another exactly when it has all the
properties of the second, and more besides.\footnote{Eder and
Ramharter mistakenly state that this is a partial order, but it is not
reflexive.  In fact, it is a total order (i.e., irreflexive and
transitive), but may be sparse or unconnected (i.e., not
trichotomous).  To see the latter point, suppose that all beings exist
in reality, but have no other properties.  Then no being is greater
than any other, but each is something ``than which there is no
greater.''}  This treatment is higher order because it involves
quantification over properties, not merely individuals.  This is seen
in the definition of \texttt{>} in the PVS formalization of Eder and
Ramharter's higher order treatment shown in Figure \ref{eandrho}.
Notice that \texttt{P} is a set (which is equivalent to a predicate in
higher-order logic) of predicates on \texttt{beings}; in PVS a
predicate in parentheses as in \texttt{F: VAR (P)} denotes the
corresponding subtype, so that \texttt{F} is a variable ranging over
the subsets of \texttt{P}.  A tutorial-level description of this PVS
formalization is provided elsewhere \citep{Rushby:ER-OntArg16}.

\begin{figure}[!ht]
\begin{session}\small
EandRho: THEORY
BEGIN

  beings: TYPE

  x, y, z: VAR beings

  re?: pred[beings]

  P: set[ pred[beings] ]

  F: VAR (P)

  >(x, y): bool = (FORALL F: F(y) => F(x))\(\,\)&\(\,\)(EXISTS F: F(x) AND NOT F(y))

  God?(x): bool = NOT EXISTS y: y > x

  ExUnd: AXIOM EXISTS x: God?(x)

  Realization: AXIOM
    FORALL (FF:setof[(P)]): EXISTS x: FORALL F: F(x) = FF(F)

  God_re_ho: THEOREM member(re?, P) => EXISTS x: God?(x) AND re?(x)

END EandRho
\end{session}
\caption{\label{eandrho}Eder and Ramharter's Higher Order Treatment, in PVS}
\end{figure}

Before examining question begging in this version, note that
\texttt{Realization} can be used to prove the following premise.
\begin{sessionlab}{PVS fragment}
  Greater_triv: LEMMA
     member(re? P) IMPLIES FORALL x: God?(x) => re?(x)
\end{sessionlab}
Given this and \texttt{ExUnd}, it is trivial to prove the
conclusion \texttt{God\_re\_ho} without opening the definition of
\texttt{God?}.  Thus, the theory with \texttt{Greater\_triv} in place
of \texttt{Realization} is vacuous and, by similar reasoning to
Section \ref{vacuity}, we argue that the original Figure \ref{eandrho}
is as well.

The strategy for proving \texttt{God\_re\_ho} in Figure \ref{eandrho}
is first to consider the being \texttt{x} introduced by
\texttt{ExUnd}; if this being exists in reality, then we are done.  If
not, then we consider a new being that has exactly the same properties
as \texttt{x}, plus existence in reality---this is attractively close
to Anselm's own strategy, which is to suppose that very same being can
be (re)considered as existing in reality.  In the PVS proof this is
accomplished by the proof step
\begin{sessionlab}{PVS Proof Step}
(name "X" "choose! z: FORALL F: F(z) = (F(x!1) OR F=re?)")
\end{sessionlab}
which names \texttt{X} to be such a being.  Here, \texttt{x!1} is the
Skolem constant corresponding to the \texttt{x} introduced by
\texttt{ExUnd} and \texttt{choose!} is a ``binder'' derived from the
PVS choice function \texttt{choose}, which is defined in the PVS
Prelude.  This \texttt{X} is some being that satisfies all the
predicates of \texttt{x!1}, plus \texttt{re?}.  Given this \texttt{X},
we can complete the proof, except that PVS generates the subsidiary
TCC proof obligation shown below to ensure that the application of the
choice function is well-defined (i.e., there is such an \texttt{X}).
\begin{sessionlab}{PVS TCC}
EXISTS (x: beings): (FORALL F: F(x) = (F(x!1) OR F = re?))
\end{sessionlab}
This proof obligation requires us to establish that there is a being
that satisfies the expression in the \texttt{choose!}; it is generated
from the predicate subtype specified for the argument to
\texttt{choose}.\footnote{This is similar to the proof obligation
generated for the definite description used in Oppenheimer and Zalta's
rendition of Figure \ref{oandz}: there, we had to prove that the
predicate in \texttt{the} is uniquely satisfiable; here we need merely
to prove that the predicate in \texttt{choose!} is satisfiable.  The
properties of the definite description, the choice function, and
Hilbert's $\varepsilon$ are described and compared in our description
of Oppenheimer and Zalta's treatment \citep{Rushby:ontological13}.}

Eder and Ramharter provide the axiom \texttt{Realization} for this
purpose; it states that for any collection of properties, there is a
being that exemplifies \emph{exactly} those properties and, when its
variable \texttt{FF} is instantiated with the term $$\texttt{\{\,G:
(P) | G(x!1) OR G=re?\,\}},$$ it provides exactly the expression
above.  In other words, \texttt{Realization} is a generalization of
the formula required to discharge a crucial step (namely, the TCC
above) in the proof.  Thus, I claim that the premise
\texttt{Realization} indirectly begs the question in this proof.  This
seems appropriate to me, because \texttt{Realization} says we can
always ``turn on'' real existence and, taken together with
\texttt{ExUnd} and the definition of \texttt{>}, this amounts to the
desired conclusion, whose ``hiding place'' (see Section \ref{compare})
is thereby revealed.

An alternative and more common style of proof in PVS would invoke the
premise \texttt{Realization} directly at the point where \texttt{name}
and \texttt{choose!} are used in the proof described above.  The direct
invocation obscures the relationship between the formal proof and
Anselm's own strategy, and it also uses \texttt{Realization} as one
step in a chain of deductions that masks its question begging
character.  Thus, use of \texttt{name} and \texttt{choose!} are key to
revealing both the strategy of the proof and the question begging
character of \texttt{Realization}.  Note that the deductions prior to
the \texttt{name} command, and those on the subsequent branch to
discharge the TCC should be routine if \texttt{Realization} is to be
considered indirectly question begging, but those on the other branch
may be arbitrarily complex.

\vspace*{-2ex}
\subsubsection*{Addendum}  If we
perform the substitution mentioned above for the variable \texttt{FF}
in \texttt{Realization} and do some deduction, then we arrive at the
following formula.   
\begin{session}
Greater_triv1: LEMMA member(re?, P) => FORALL x: re?(x) OR EXISTS y: y>x
\end{session}
The right hand disjunct is just \texttt{NOT God?(x)}, and then some
propositional rearrangement gives us the following.
\begin{session}
Greater_triv2: LEMMA member(re?, P) => FORALL x: God?(x) => re?(x)
\end{session}
But this clearly begs the question: it takes us straight from
\texttt{ExUnd} to the conclusion \texttt{God\_re\_ho}.  Since
\texttt{Realization} entails this formula, we have an intuitive reason
why it also should be considered question begging.  However, proof of
entailment uses nonobvious quantifier reasoning (the instantiation for
\texttt{FF} mentioned above) and we could cite this to exonerate
Realization of begging the question.  \textbf{End of addendum}\\[-1ex]

\cite{Campbell18:book} adopts some of Eder and Ramharter's
higher order treatment, but rejects \texttt{Realization} on the
grounds that it is false.  Observe that we could have incompatible
properties\footnote{Eder and Ramharter are careful to require that all
the greater-making properties are ``positive'' so directly
contradictory properties are excluded, but we can have positive
properties that are mutually incompatible
\citep{iep-ontological-argument}.  Examples are being ``perfectly
just'' and ``perfectly merciful'': the first entails delivering
exactly the ``right amount'' of punishment, while the latter may
deliver less than is deserved.}  and \texttt{Realization} would then
provide the existence (in the understanding) of a being that
exemplifies those incompatible properties, and this is certainly
questionable.  A better approach might be to weaken
\texttt{Realization} to allow merely the addition of \texttt{re?}\ to
the properties of some existing being.  This is essentially the
approach taken below.

\begin{figure}[!ht]
\begin{session}\small
Campbell: THEORY
BEGIN

  beings: TYPE

  x, y, z: VAR beings

  re?: pred[beings]

  P: set[ pred[beings] ]

  F: var (P)

  >(x, y): bool = (FORALL F: F(y) => F(x))\(\,\)&\(\,\)(EXISTS F: F(x) AND NOT F(y))

  God?(x): bool = NOT EXISTS y: y > x

  ExUnd: AXIOM EXISTS x: God?(x)

  quasi_id(D: setof[(P)])(x,y: beings): bool = 
     FORALL (F:(P)): NOT D(F) => F(x) = F(y)

  jre: setof[(P)] = singleton(re?) 

  Weak_real: AXIOM
      NOT re?(x) => (EXISTS z: quasi_id(jre)(z, x) AND re?(z))

  God_re_ho: THEOREM member(re?, P) => EXISTS x: God?(x) AND re?(x)

END Campbell
\end{session}
\caption{\label{campbell}Simplified Version of Campbell's Treatment, in PVS}
\end{figure}

Campbell's full treatment \citep{Campbell18:book} differs from others
considered here in that he includes more of Anselm's presentation of
the Argument (e.g., where he speaks of ``the Fool'').  The treatment
shown in Figure \ref{campbell} is my simplified interpretation of
Campbell's approach, scaled back to resemble the other treatments
considered here, and is based on discussions prior to publication of
his book \citep{Campbell16}.  Campbell adopts Eder and Ramharter's
higher order treatment, but replaces \texttt{Realization} by (in my
interpretation) the axiom \texttt{Weak\_real} which essentially states
that if \texttt{x} does not exist in reality, then we can consider a
being just like it that does.  A being ``just like it'' is defined in
terms of a predicate \texttt{quasi\_id} introduced by \citet[Section
3.3]{Eder&Ramharter15} and is true of two beings if they have the same
properties, except possibly those in a given set \texttt{D}\@.
Observe that the PVS specification writes this higher order predicate
in Curried form.  Here, \texttt{D} is always instantiated by the
singleton set \texttt{jre} containing just \texttt{re?}, so we always
use \texttt{quasi\_id(jre)}.

A couple of routine proof steps bring us to the following sequent.
\begin{session}
God_re_ho :  

\{-1\}  P(re?)
  |-------
[1]   EXISTS y: y > x!1
[2]   re?(x!1)

\end{session}
Our technique for discharging this is to instantiate formula 1 with a
being just like \texttt{x!1} that does exist in reality, which we name
\texttt{X}.
\begin{sessionlab}{PVS Proof Step}
(name "X" "(choose! z: quasi_id(jre)(z, x!1) AND re?(z))")
\end{sessionlab}
The main branch of the proof then easily completes and we are left
with the TCC obligation to ensure that application of the choice
function is well-defined.   That is, we need to show
\begin{sessionlab}{PVS TCC}
EXISTS (z: beings): quasi_id(jre)(z, x!1) AND re?(z)
\end{sessionlab}
under the condition \texttt{NOT re?(x!1)}.   This is precisely what
the premise \texttt{Weak\_real} supplies, so we may conclude that this
premise indirectly begs the question.

We should also note that \texttt{Weak\_real} can be used to prove
\texttt{Greater\_triv}, as discussed for Figure \ref{eandrho} at the
beginning of this section, and is therefore vulnerable to the charge
of vacuity.

\vspace*{-2ex}
\subsubsection*{Addendum} The premise \texttt{Weak\_real} entails the
obviously question begging formula \texttt{Greater\_triv2}, introduced
in the previous addendum, just as \texttt{Realization} does.  However,
unlike the previous case, only trivial quantifier reasoning is used and
thereby provides an intuitive reason why \texttt{Weak\_real} should be
considered to beg the question.  \textbf{End of addendum}\\[-1ex]

The higher order formalizations considered in this section have
slightly longer and more complex proofs than those considered earlier.
This means that the indirect question begging character of a
particular premise may not be obvious if it occurs in the middle of a
chain of proof steps.  Use of the \texttt{name} and \texttt{choose!}
constructs accomplishes two things: it highlights the strategy of the
proof (namely, it identifies the attributes of the alternative being
to consider if the first one does not exist in reality), and it
isolates application of the questionable premise to a context where
its indirect question begging character is revealed.

\section{Comparison with Informal Accounts of\newline Begging the Question}
\label{compare}

There are several works that examine the Ontological Argument against
the charge that it begs the question.  Some of them, including this
one, employ a ``logical'' interpretation for begging the question,
which is to say they associate question begging with the logical form
of the argument and not with the meaning attached to its symbols.
Others employ a ``semantical'' interpretation and find circularities
in the meanings of the concepts employed by the Argument prior to
consideration of its logical form.

\cite{Roth70}, for example, observes that Anselm begins by
offering a definition of God as that than which nothing greater can be
conceived; Roth then claims that greatness already presupposes
existence and is therefore question begging.   \cite{McGrath90}
criticizes Roth's analysis and presents his own, which finds
circularity in the relationship between possible and real existence.
 \cite{Devine75} (who was writing 15 years earlier than McGrath
but is not cited by him) asks whether it is possible to use ``God'' in
a true sentence without assuming His existence and concludes that it
is indeed possible and thereby acquits the Argument of this kind of
circularity.

All these considerations lie outside the scope considered here.  We
treat ``greater than,'' ``real existence,'' and any other required
terms as uninterpreted constants, and we assume there is no conflict
between the parts they play in the formalized Argument and the
intuitive interpretations attached to them.  We then ask whether the
formalized argument begs the question in a logical sense.

Many authors consider logical question begging in semi-formal
arguments.  Some consider a ``dialectical'' interpretation associated
with the back and forth style of argumentation that dates to
Aristotle's original identification of the fallacy (as he thought of
it), while others consider an ``epistemic'' interpretation in the
context of standard deductive arguments.  \cite{Walton94} outlines a
history of analysis of begging the question, focusing on the
dialectical interpretation, while \cite{Garbacz02} provides a formal
account within this framework.  \cite{Walton06} contends that the
notion of question begging and the intellectual tools to detect it are
similar in both the dialectical and epistemic interpretations, so I
will focus on the epistemic case.  The intuitive idea is that a
premise begs the question epistemically when ``the arguer's belief in
the premise is dependent on his or her reason to believe the
conclusion'' \cite[page 241]{Walton06}.

Several authors propose concrete definitions or methods for detecting
epistemic question begging.   \cite{Walton06}, for example,
recommends proof diagrams (as supported in the Araucaria system
\citep{Reed04:araucaria}) as a tool to represent the structure of
informal arguments, and hence reveal question begging circularities.
He illustrates this with ``The Bank Manager Example'':
\begin{description}\itemsep=0ex
\item[Manager:] Can you give me a credit reference?
\item[Smith:] My friend Jones will vouch for me.
\item[Manager:] How do we know he can be trusted?
\item[Smith:] Oh, I assure you he can.
\end{description}

Our interest here is with formal arguments and as soon as one starts
to formalize The Bank Manager Example, it becomes clear that the
argument is invalid, for it has the following form.
\begin{description}\itemsep=0ex 
\item[Premise 1:] $\forall a, b: \texttt{trusted}(a) \wedge
\texttt{vouch-for}(a, b) \supset \texttt{trusted}(b)$
\item[Premise 2:] $\texttt{vouch-for}(\texttt{Jones}, \texttt{Smith})$
\item[Premise 3:] $\texttt{vouch-for}(\texttt{Smith}, \texttt{Jones})$
\item[Conclusion:] $\texttt{trusted}(\texttt{Smith})$
\end{description}
The invalidity here is stark and independent of any ideas about
question begging.  Walton describes other methods for detecting
question begging in informal arguments but most of the examples are
revealed as invalid when formalized.  While these methods may be of
assistance to those committed to notions of informal argument or
argumentation, our focus here is on valid formal arguments, so we
do not find these specific techniques useful, although we do subscribe
to the general ``epistemic'' model of question begging, and will
return to this later.

\cite{Barker78}, building on \cite{Barker76} and \cite{Sanford77},
calls a deductive argument \emph{simplistic} if it has a premise that
entails the conclusion; he claims that all and only such (valid)
arguments are question begging.  Our definition for strict begging
includes this case, but also others.  For example, Barker considers
the argument with premises $p$ and $\neg q$ and conclusion $p$ to beg
the question, whereas that with premises $p \vee q$ and $\neg q$, and
the same conclusion does not, which seems peculiar to say the least.
Both of these are question begging by our strict definition.

Now one might try to ``mask'' the question begging character of an argument
that satisfies Barker's definition by adding obfuscating material, so
he needs some notion of equivalence to expose such ``masked''
arguments.  However, it cannot be logical equivalence of the premises
because the conjunction of premises is identical in the two cases
above, yet Barker considers one to be question begging and the other
not.  Barker proposes that ``relevant equivalence'' (i.e., the
bidirectional implication of relevance logic \citep{Dunn86}) of the
premises is the appropriate notion.  The examples above are not
equivalent by this criterion ($\neg q \supset p$ and $\neg q$
illustrate premises that are equivalent to the second example by this
criterion) and so the question begging character of the first does not
implicate the second, according to Barker.

As noted, all these examples strictly beg the question by my
definition and I claim this is as it should be.  Recall that a premise
strictly begs the question when it is equivalent to the conclusion,
given the other premises.  Now, the essence of the epistemic
interpretation for begging the question is that truth of the premise
in question is difficult to know or believe independently of the
conclusion, and I assert that this judgment must be made after we have
digested the other premises (otherwise, what is their purpose?).
Thus, if $\neg q$ is given (digested), then $p \vee q$ and $p$ are
logically equivalent and we cannot believe one independently of the
other and $p \vee q$ is rightly considered to beg the question in this
context.  Barker judges $p \vee q$ and $p$ in the absence of any other
premise and thereby reaches the wrong conclusion, in my opinion.

My proposal for strict begging differs from those in the literature
but is not unrelated to proposals such as Barker's.  However, my
proposals for weak and indirect begging depart more radically from
previous treatments.  I consider a premise to be weakly begging when
light augmentation to the other premises render it strictly begging.
Human judgment must determine whether the augmentation required is
innocuous or contrived and this can be guided by epistemic
considerations: if the augmentation is required to establish a context
in which the questionable premise(s) are plausible (as in our example
of Figure \ref{eandr1}, where we certainly intend the $>$ relation to
be nonempty), then the questionable premise(s) surely beg the question
in the informal epistemic sense as well as in our formal weak sense.

Indirect begging arises when the questionable premise supplies (a
generalization of) exactly what is required to make a key move in the
proof.  Provided we have not applied anything beyond routine
deduction, I claim that the proof state (conveniently represented as a
sequent) represents our epistemic state after digesting the other
premises and the desired conclusion.  An indirectly begging premise is
typically (a generalization of) one that can be reverse engineered
from this state, and belief in such a premise cannot be independent of
belief in the current proof state; hence such a premise begs the
question in the informal epistemic sense as well as in our formal
indirect sense.

Most authors who examine question begging in the Ontological Argument
implicitly apply an epistemic criterion, and do so in the context of
modal representations of the argument (which I examine elsewhere
\citep{Rushby:modalont19}).  \cite{Walton78}, however, does discuss first-order
formulations in a paper that is otherwise about modal formulations.

Walton begins with a formulation that is identical (modulo notation)
to that of Figure \ref{eandr2}.  He asserts that the premise
\texttt{Greater2} (his premise 2) is implausibly strong because it
``would appear to imply, for example, that a speck of dust is greater
than Paul Bunyan.''\footnote{Paul Bunyan is a lumberjack character in
American folklore.}  I would suggest that a better indicator of its
``implausible strength'' is the fact that it indirectly begs the
question, as described in Section \ref{indirect}.  Walton then
proposes that premise \texttt{Greater1} of Figure \ref{eandr1} (his
premise 2G) may be preferable but worries that our reason for
believing \texttt{Greater1} must be something like \texttt{Greater2}.
It is interesting that Walton does not indicate concern that
\texttt{Greater1} might beg the question, whereas our analysis shows
that it is weakly begging, and becomes strictly so in the presence of
premises that require a modicum of connectivity in the \texttt{>}
relation (recall Sections \ref{direct} and \ref{weak}).  Thus, I
suggest that the formulations and methods of analysis proposed here
are more precise, informative, and checkable than Walton's and other
informal interpretations for begging the question.

My three criteria for begging the question---strict, weak, and
indirect---identify question begging premises in a fairly unequivocal
manner.\footnote{Weak begging admits some equivocation in the choice
of augmenting premises, and indirect begging in the amount of
deductive effort expended.}  They provide formal interpretations for
the informal notion of ``epistemic'' begging, but it is not immediate
from either this derivation or their own definitions that these kinds
of question begging should be considered defects.  
\citet[Section 1.2(5)]{Eder&Ramharter15} observe that the conclusion to
a deductively valid argument must be implicit or ``contained'' in the
premises (otherwise, the reasoning would not be deductive), but an
argument can only be persuasive or interesting ``if it is possible to
accept the premises without already recognizing that the conclusion
follows from them.  Thus, the desired conclusion has to be `hidden' in
the premises.''  In a footnote, they aver ``Sometimes, proofs of the
existence of God are accused of being question-begging, but this
critique is untenable.  It is odd to ask for a deductive argument
whose conclusion is not contained in the premises.  Logic cannot pull
a rabbit out of the hat.''

Eder and Ramharter are, of course, correct that the conclusion must be
``contained'' in the premises, but they are also correct that it
should be ``hidden,'' and so I challenge their claim that accusations
of question begging are untenable.  I suggest that tests for question
begging should expose the ``hiding place'' of the conclusion among the
premises: if this is revealed as inadequate or contrived, then our
interest in the argument, and its persuasiveness, are diminished.

I claim that my criteria perform this function.  Strict begging tells
us that the conclusion is barely hidden at all.  However, a legitimate
criticism of strict begging is that because it applies after we have
accepted the other premises, it can indict premises that are merely
there ``to get the argument off the ground''; typically these are
simple existential premises that provide Skolem constants.  Examples
are \texttt{ExUnd}, which appears in all the formalizations
considered, and \texttt{Ex\_re}, which appears in Figure \ref{eandr2}.
Human judgment exonerates these.  However, strict begging does
correctly indict the premise \texttt{Greater1} in Figure \ref{oandz}
and weak begging does the same in Figure \ref{eandr1}.  This premise
obfuscates the obviously question-begging premise
\texttt{Greater1\_vac} of Figure \ref{vac1} by expanding the
definition of \texttt{God?}  and using the contrapositive, and strict
and weak begging expose this subterfuge.

Indirect begging identifies premises that are equivalent to those that
would be constructed by reverse-engineering from the conclusion and
other premises: to my mind, it reveals contrivance.  One reason for
the enduring interest in the Ontological Argument is surely that its
premises seem innocuous, yet its conclusion is bold.  But when a
premise is revealed as indirectly begging, we see how the ``trick'' is
performed and this must eliminate our surprise and diminish our
delight.

In summary, my criteria for begging the question are consistent with,
and give formal expression to, informal ``epistemic'' interpretations.
That is, a question begging premise is one that is difficult to
understand or believe independently of the conclusion.  In addition,
my criteria expose premises that ``hide'' the conclusion or a key step
in its proof in ways that can suggest contrivance or obfuscation.

\subsubsection*{Addendum} \cite{Oppenheimer&Zalta21} criticize my
definitions for begging the question because this ``fallacy\ldots
traditionally applies to \emph{arguments}, not to \emph{premises}, yet
for some reason Rushby takes it to apply to premises.''  Contrary to
their claim, begging the question is universally applied, not to a
whole argument, but to specific premises within the context of an
argument, as can be seen in the discussion and works cited earlier in
this section.  In the case of the Ontological Argument, we can look to
other papers that specifically raise the issue of begging the
question.  For example, in his highly cited work \cite{Lewis70}
devotes considerable attention to his Premise 3 and declares version
3C to be the most attractive variant but worries that is it
``circular.''  \cite{Rowe76} claims that the traditional Ontological
Argument begs the question.  This is challenged by
\cite{Davis76-begging}, in part because Rowe does not specify what he
means by begging the question; \cite{Rowe76-comments} clarifies this
and \cite{Davis76-reply}, still unpersuaded, responds, and is followed
by others \citep{Wainwright78,Loptson80:AnselmMeinong,Walton78}; all
these authors focus on specific premises within Rowe's formulation of
the argument.\footnote{I have formally analyzed Lewis' and Rowe's
treatments \citep{Rushby:modalont19}.}  

My own definition for strict begging follows exactly this pattern: it
concerns the relationship between the questionable premises and the
conclusion in the context given by the other premises (i.e., the whole
argument).  My other definitions have the same character.  When $C$ is
our conclusion, $Q$ our ``questionable'' premise(s) and $P$ our other
premises, we have $Q, P \vdash C$ and must therefore have $Q \supset
(P \supset C)$ (by the deduction theorem).  If $Q$ states directly $P
\supset C$, then I call it ``\emph{obviously} question begging'' since
there is no hope of the argument yielding surprise or delight.  $Q$
must therefore hide or obfuscate this relationship; typically it does
so by taking its contrapositive, folding/unfolding some definitions
and, sometimes, using quantification to strengthen the formula.  My
definitions reveal such premises and label them ``question begging''
but the extent to which they ``truly'' beg the question depends on how
we interpret the obfuscation.  If we think the obfuscated premise was
reverse-engineered to fulfill its role in the argument, then it seems
fair to say it begs the question, but if we think the author proposed
it for conceptual reasons, then we can exonerate it.

\cite{Oppenheimer&Zalta21} continue their criticism with an example
that has premises $A \equiv B$, $B \equiv C$ and conclusion $C \equiv
A$; they assert this is non-question begging, but that if we add a
third premise $A$ and change the conclusion to just $C$ then I would
accuse the modified argument of question begging even though it is
equivalent to the first.  If their point is that my definition is
inconsistent, then they are wrong, for it also identifies the premise
$B \equiv C$ (or $A \equiv B$) as question begging in the first
argument.  If their point is that my definitions are insufficiently
discriminating, then I accept the charge.  I have already noted that
the premises \texttt{ExUnd} and \texttt{Ex\_re} are unreasonably
accused of strict begging in Sections \ref{weak} and \ref{indirect};
my methods suggest locations where question begging may be hiding and
how it is hidden, but they are not unequivocal and it is the reader's
choice whether or not they find them interesting.  \textbf{End of
addendum}\\[-1ex]

\section{Conclusion} \label{conc}

Once we go beyond the ``simplistic'' case \citep{Barker78}, where the
conclusion is directly entailed by one of its premises, the idea of
begging the question is open to discussion and personal judgment.  A
variety of positions are contested in the literature on argumentation
and were surveyed in Section \ref{compare}, but I have not seen any
discussion of question begging in fully formal deductive settings.

My proposal is that a premise may be considered to beg the question
when it is equivalent to the conclusion, given the other premises
(strict begging), or a light augmentation of these (weak begging), or
when it directly discharges a key step of the proof (indirect
begging).  The intuition is that such premises are so close to the
conclusion or its proof that they cannot be understood or believed
independently of it, and their construction seems
``reverse-engineered'' or otherwise contrived so that surprise and
interest in the argument is diminished.

I have shown that several first- and higher-order formalizations of
the Ontological Argument beg the question, illustrating each of the
three kinds of question begging.  I suspect that all similar
formulations of the Argument are vulnerable to the same charge.
Separately (in work performed after that described here), I have
examined several formulations of the argument in quantified modal
logic (including that of  \cite{Rowe76}, who explicitly accuses
the Argument of begging the question, and those of 
\cite{Adams71} and \cite{Lewis70}, who also discuss circularity)
and found them vulnerable to the same criticism
\citep{Rushby:modalont19}.

Begging the question is not a fatal defect and does not affect
validity of its argument; identification of a question begging premise
can be an interesting observation in its own right, as may be
identification of the augmented premises that reveal a weakly begging
one.  However, I think most would agree that the persuasiveness of an
argument is diminished when its premises are shown to beg the
question.  Furthermore, revelation of question begging undermines any
delight or surprise in the conclusion, for the question begging
premise is now seen to express the same idea.
Indirect begging is perhaps the most delicate case: it
reveals how exquisitely crafted---one is always tempted to say
reverse-engineered---is the questionable premise to its r\^{o}le in
the proof.  To my mind, it casts doubt on the extent to which the
premise may be considered analytic in the sense that 
\citet[Section 1.2(7)]{Eder&Ramharter15}
use the term: that is, something that the author ``could
have held to be true for conceptual (non-empirical) reasons.''

I have also shown that all the formalizations of the argument
considered here entail variants that are vacuous: that is, apply no
interpretation to ``than which there is no greater'' (formally, they
leave the predicate \texttt{God?} uninterpreted).  I think this a more
serious and overarching defect than question begging.  To see this,
observe that all the formalizations considered here can be
reconstructed by the following procedure.  The first four steps
construct a vacuous formalization similar to Figure \ref{vac1}.

\begin{enumerate}
\item Introduce uninterpreted predicates \texttt{God?} and
\texttt{re?} over \texttt{beings}.

\item Introduce a premise similar to \texttt{ExUnd} and, optionally,
one similar to \texttt{Ex\_re} (Figure \ref{eandr2}).

\item Specify a conclusion, similar to \texttt{God\_re\_alt}:
some \texttt{being} satisfies both \texttt{God?} and \texttt{re?}.

\item Reverse-engineer a premise \texttt{Greater} that entails the
conclusion, given the other premises: something like
\texttt{Greater1\_vac} or \texttt{Greater2\_triv} (Section
\ref{indirect}).
\end{enumerate}
At this point, we have a reconstruction similar to Figure \ref{vac1}
that is valid but vacuous, because there is no interpretation for
\texttt{God?}.  Hence, it remains valid when any interpretation is
supplied, including Gaunilo's most perfect island.

The reverse-engineered premise \texttt{Greater} indirectly begs the
question because (due to its method of construction) it cannot be
understood or believed independently of the conclusion and the other
premises (so it is question begging ``in the epistemic sense'').

\begin{enumerate}\setcounter{enumi}{4}

\item Supply an interpretation (i.e., a definition) for \texttt{God?}
and replace some or all appearances of this predicate by its
definition.  The interpretation may require additional terms and
premises, such as an interpreted or uninterpreted \texttt{>} relation,
and higher-order constructions.

\item Optionally, adjust the resulting reconstruction (e.g., by
propositional rearrangement, by adding terms, or by adding variables
and adjusting quantification), taking care that it remains valid
(typically an adjusted premise will entail the original).

\end{enumerate}
The reconstruction following step 4 is vacuous and begs the question
and I maintain that the adjustments made in steps 5 and 6 cannot
remove these characteristics (although they can obfuscate them) and 
we cannot attach any belief to the premises once we see how they are
constructed. 

Sections \ref{vacuity} through \ref{ho} provided evidence that all the
formalized arguments examined there can be reconstructed by this
procedure.  I do not claim the original authors of those formalized
arguments took this route---they were surely sincere and constructed
premises that they considered both analytic and faithful to Anselm's
intent---but its existence exposes the hollow nature of all these
formalizations.

It is, of course, for individual readers to form their own opinions
and to decide whether the forms of question begging and vacuity
identified here affect their confidence, or their interest, in the
various renditions of Anselm's Argument, or in the Argument itself.
What I hope all readers find attractive is that these methods provide
explicit evidence to support accusations of question begging that can
be exhibited, examined, and discussed, and that may be found
interesting or enlightening even if the accusations are ultimately
rejected.

Observe that detection of the various kinds of question begging
requires exploring variations on a specification or proof.  This is
tedious and error-prone to do by hand, but simple, fast, and reliable
using mechanized assistance.  I hope the methods and tools illustrated
here will encourage others to investigate similar questions concerning
this and other formalized arguments: as Leibniz said, ``let us
calculate.''

\subsection*{Acknowledgments}

I am grateful to Richard Campbell of the Australian National
University for stimulating and dissenting discussion on these topics,
to my colleagues Sam Owre and N.\ Shankar for many useful
conversations on PVS and logic, and to anonymous reviewers of this and
a previous version for very helpful comments

\bibliographystyle{spbasic}

\end{document}